\documentstyle[epsfig,pre,multicol,aps]{revtex}

\begin{document}
\begin{title}
{\bf Comment on the paper 'Thermodynamics of two-dimensional magneto
nanoparticles' by P. Vargas, D. Altbir, M. Knobel and D. Laroze } 
\end{title}
\author{H. B{\"u}ttner, Yu. Gaididei}
\address{Physikalishes Institut, Universit\" at Bayreuth, 
D95440 Bayreuth, Germany\\
Bogolyubov Institute for Theoretical Physics, Kiev, Ukraine}

\maketitle

\bigskip 
\bigskip

\begin{abstract}It is shown that there is no bi-stability in the system of
two-dimensional nanoparticles considered by  
P. Vargas et al. in Ref. \cite{vargas}.
\end{abstract}

In the recent Letter \cite{vargas}  the thermodynamics of two-dimensional
magnetic nanoparticles in an external magnetic field has been considered and
equivalent mechanical model was presented.
We want to comment on their    result
that the equilibrium magnetization has a maximum at finite
temperatures and on their  mechanical  analogue result.

We start with the mechanical model which is presented in Fig. 1 of
\cite{vargas}. This model is
desribed by the following  three dimensional Lagrangian function
\begin{eqnarray}\label{lagr}
{\cal L}=\frac{m}{2}\,\left(\dot{x}^2+\dot{y}^2+\dot{z}^2\right)+m \omega
\,\left(\dot{x} \,y-x\,\dot{y}\right)+mgz \end{eqnarray}
with a constraint \begin{eqnarray}\label{constr}x^2+y^2+z^2=R^2\end{eqnarray}
where $m$ is the mass of the particle, $R$ is the radius of the ring, $g$ is
the gravity acceleration. The second term in Eq. (\ref{lagr}) represents an
inertia contribution due to the rotation of the ring around the $z$-axis. By
using the Legendre transformation \cite{landau} we get the Hamiltonian of the
system in the form \begin{eqnarray}\label{hamil}
H=\frac{1}{2m}\left(p_x^2+p_y^2+p_z^2\right)-\omega
\,\left(x\,p_y-y\,p_x\right)+mgz \end{eqnarray}
From Eq. (\ref{hamil}) we obtain  the partition function in the general form
$$Z_{mech}=N\,\int\limits_{-\infty}^{\infty}\,dp_x\,dp_y\,dp_z
\int\limits_{-\infty}^{\infty}\,d_x\,d_y\,d_z\,e^{-H/kT}\,\delta\left(\sqrt{x^2+y^2+z^2}-R\right)$$
where $N$ is  a normalization factor, $k$ is the Boltzman constant and the
constraint (\ref{constr}) was taken into account. This can be written for (\ref{hamil}) as
\begin{eqnarray}\label{part}
Z_{mech}=2\pi R^2
N\int\limits_{0}^{\pi}\,d\theta\,\sin\theta\,\exp\{-mgR\frac{{\cal U}}{kT}\}
\end{eqnarray}
with $${\cal U}=-\omega^2 R^2/g \sin^2 \theta-\cos\theta$$ where the angle
$\theta$ is measured with respect to the z-axis. This
potential function coincides with the  potential  function
introduced in \cite{vargas} but  due to the condition that the angle $\theta$
varies only in the interval $(0,\pi)$ there is {\em no} bistability in the system.
Another point which is missing in the partition function $Z_{2D}$ presented in
\cite{vargas} is the factor $\sin\theta$ in the Jacobian of the transformation
from the cartesian- to the polar co-ordinates.

Thus we  see
that the interpretation of the dimensionless energy ( see Fig. 2 of
\cite{vargas}) is misleading because
it does not present a double well potential and therefore there is no bi-
stability of the system. This can  easily be seen from the fact that
the two angles for minimum energy belong to two different directions of rotation:
 the positive angle  for the anti-clockwise, the negative angle for the clock-wise
rotation. With
this remarks in mind it should be also obvious that there is
 no bifurcation for a given rotation ( Fig. 3 in
\cite{vargas}).

When Vargas et al. \cite{vargas} analyze the thermodynamical properties of magnetic 
nanoparticles  they used  the enrgy of the single particle in the form
\begin{equation} \label{vargen}
E=-m H\cos\theta-\frac{1}{2} m H_a \sin^2\theta\end{equation} 
where $H$ is an external magnetic field and $H_a$ is the anisotropy field.
It is clearly seen  that in the absence of magnetic field $H$  Eq. (\ref{vargen})
gives the energy  of an {\em  easy-plane} magnet : the magnetic moment can
freely rotate  in the plane perpendicular  to the z-axis. Note in
passing that this  is in a complete agreement with the above-considered mechanical 
model  when in the no-gravity case the particle may be situated at any point of
the circular ring of  radius $R$. We would like to remark that the desription of
an easy-axes magnet is quite another problem and does not correpond to the above
mechanical model.

\begin{figure}
\centerline{\hbox{
\epsfig{figure=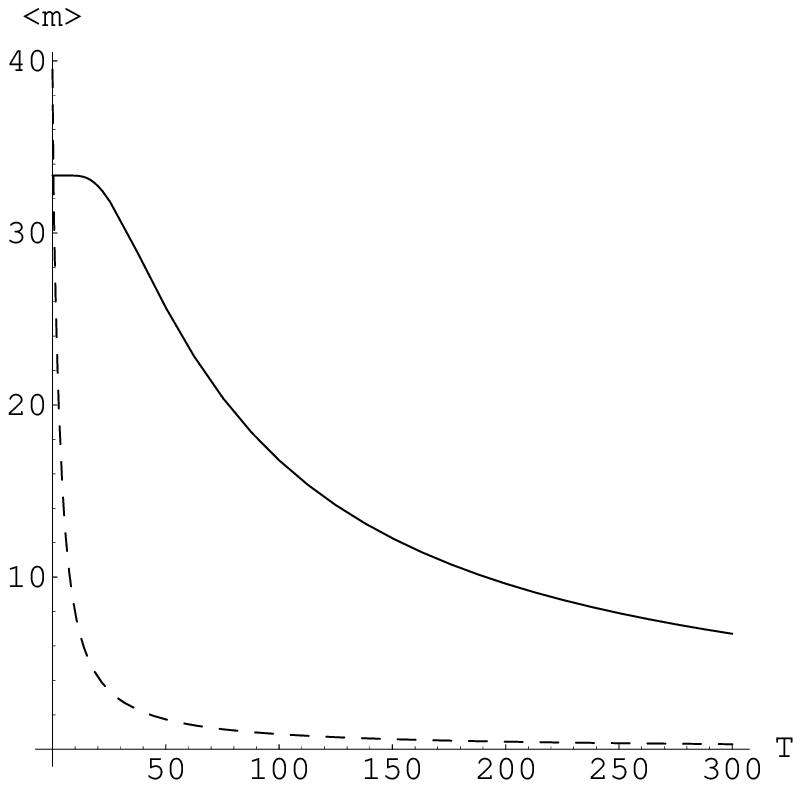,width=70mm,angle=0}
\epsfig{figure=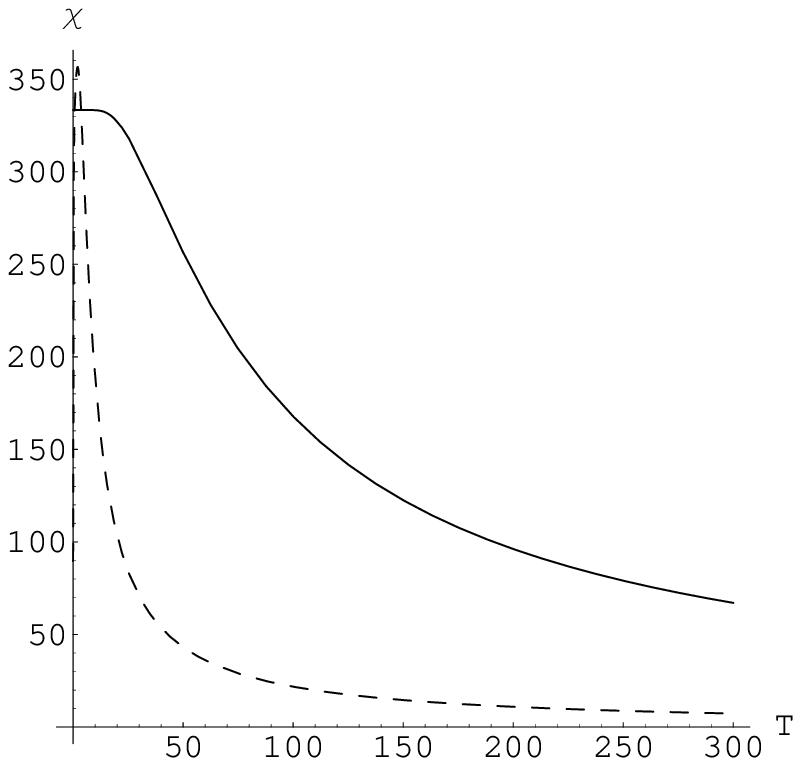,width=70mm,angle=0}}}
\caption{ Magnetization in units of $\mu_B$ (left figure) and
magnetic susceptibility in units $\mu_b/kOe$  (right figure), as a function of
temperature for the same set of parameters as in Ref.[1]. The solid line
corresponds to $H=0.1kOe,~H_a=3 kOe$, the dashed line gives the results for
$H=0.1kOe,~H_a=0.09 kOe$. In the latter case the magnetization is rescaled by
the factor 0.05 while the susceptibility is rescaled by the factor 0.1} 
\label{fig:Fig1} \end{figure}

Exactly the same partition function as given in Eq. (\ref{part})  ( with
appropriate change of variables)  can be obtained for a magnetic particle
which is described by equation (\ref{vargen}).  The reason
 for the pronounced discrepancy between their partition function and that in Eq.
(\ref{part})  is  the following. The authors of \cite{vargas} overlooked the fact
that  by calculating the partition function one should integrate over the
phase space $d\Gamma=dp\,dq$ where $p$ and $q$ are canonically conjugated
variables (see e.g. \cite{landaust}). In  magnetic systems which are
characterized by the classical three dimensional magnetization vectors
$\vec{M}=\,\left(\sin\theta\cos\phi,\sin\theta\sin\phi,\cos\theta\right)$  
the $z$-component of the magnetization, $m=\cos\theta$, and the azimuthal
angle, $\phi$, are canonically conjugated variables. Thus  the element of the
phase space volume is given by $d\Gamma=\sin\theta d\theta\,d\phi$ 
and  the partition function  for magnetic particles takes the form
(\ref{part}). When one calculates the mean magnetization, $\langle m\rangle$,
and the susceptibility,  $\chi$,  along the field  by using the
correct partition function of the form
\begin{eqnarray}\label{partm}
Z_{magn}=2\pi
N\int\limits_{0}^{\pi}\,d\theta\,\sin\theta\,\exp\{\frac{mH\cos\theta+m
H_a\sin^2\theta/2}{kT}\} \end{eqnarray}
one obtains dependences shown in  Fig.\ref{fig:Fig1}.
\vspace*{20mm}

Thus in contrast to the claim of Ref.\cite{vargas} the correctly
 defined  magnetization  does not manifest non-monotonic
behaviour  as a function of temperature. The magnetic
susceptibility  has a maximum for $H_a < H$ ( not seen in Fig.1 )in the low temperature
interval but for $H_a\geq H$  it is a monotonic function of $T$.

{\bf Acknowledgements}
Yu. G. is grateful for the hospitality of the University of Bayreuth
 where this work was performed.

\end{document}